\documentclass[notitlepage,aps,pra,twocolumn,groupaddress,superscriptaddress]{revtex4-1}

\usepackage[left=2cm,top=2cm,right=2cm,headsep=0.5cm,headheight=0.5cm,nofoot]{geometry}

\usepackage[utf8]{inputenc}
\usepackage[table]{xcolor}
\usepackage{colortbl}
\usepackage{indentfirst}
\usepackage{stmaryrd}
\usepackage{amssymb,amsmath,amsthm,amsfonts,amsbsy}
\usepackage{bm,bibunits,color,chngcntr,epsfig,epstopdf,graphicx,dsfont}
\usepackage{hyperref,lipsum,,makecell,mathrsfs,rotating}
\usepackage[english]{babel}
\usepackage[normalem]{ulem}

\newcommand{\be}{\begin{equation}}
\newcommand{\ee}{\end{equation}}
\newcommand{\bea}{\begin{eqnarray}}
\newcommand{\eea}{\end{eqnarray}}
\newcommand{\ket}{\rangle}
\newcommand{\bra}{\langle}

\newcommand{\I}{\mathds{1}}
\newcommand{\ra}{\rightarrow}

\def\C#1{\mathcal #1}

\def\S#1{\mathscr #1}

\definecolor{gray}{gray}{0.9}

\begin{document}
\newtheorem{theorem}{Theorem}
\newtheorem{prop}[theorem]{Proposition}
\newtheorem{corollary}[theorem]{Corollary}
\newtheorem{open problem}[theorem]{Open Problem}
\newtheorem{conjecture}[theorem]{Conjecture}
\newtheorem{definition}{Definition}
\newtheorem{remark}{Remark}
\newtheorem{example}{Example}
\newtheorem{task}{Task}

\title{Quantum circuit simulation of superchannels}
\author{Kai Wang}
\affiliation{CAS Key Laboratory of Theoretical Physics, Institute of Theoretical Physics,
Chinese Academy of Sciences, Beijing 100190, China}
\affiliation{School of Physical Sciences, University of 
Chinese Academy of Sciences, Beijing 100049, China}
\author{Dong-Sheng Wang}\thanks{wds@itp.ac.cn}
\affiliation{CAS Key Laboratory of Theoretical Physics, Institute of Theoretical Physics,
Chinese Academy of Sciences, Beijing 100190, China}
\date{\today}



\begin{abstract}
Quantum simulation is one of the central discipline to demonstrate the
power of quantum computing. 
In recent years, the theoretical framework of quantum superchannels
has been developed and applied widely as the extension of quantum channels. 
In this work, we study the quantum circuit simulation task of superchannels.
We develop a quantum superchannel simulation algorithm based on the convex decomposition 
into sum of extreme superchannels.
We demonstrate the algorithm by numerical simulation of qubit superchannels with high accuracy,
making it applicable to current experimental platforms.
Our study stands as an expansion of the superchannel theory 
to the field of quantum simulation and algorithm, 
as well as an extension of quantum simulation from channels and open-system dynamics
to superchannels and processes with manifest quantum memory effects. 
\end{abstract}

\maketitle

\section{Introduction}

In modern quantum physics, 
the control and engineering of complex quantum dynamics is essential.
In recent years,
quantum computing has become a powerful paradigm to achieve this.
With the standard quantum circuit model, 
the field of digital quantum simulation is expected to 
be powerful to realize and study quantum dynamics.


A quantum circuit contains a sequence of gates realized at discrete time slots.
The original and still an important task is quantum Hamiltonian simulation,
with each gate in a circuit usually described by a local Hamiltonian term~\cite{Fey82}.
At the same time,
quantum circuit simulation relies on quantum gates directly and no explicit Hamiltonian.
It is more flexible and of digital nature, 
and applies to broader settings such as quantum optical systems. 

Quantum evolution is characterized by CPTP maps~\cite{Kra83,Cho75},
or called quantum channels.
Unitary evolution, Lindblad master equation,
and positive operator-valued
measures (POVM) can all be viewed as quantum channels~\cite{NC00}.
Quantum circuit simulation of quantum channels, or quantum channel simulation in short,
have been studied in the past.
These include the simulation schemes for 
quantum channel and open-system dynamics~\cite{BBW07,PPK+11,WBOS13,WS15,Wang16,SSP14,SSBP15,TV17},
dissipative quantum computing and state preparation~\cite{VWC09},  
and application in quantum thermometry~\cite{TFS+16,CMP+18},
and also various
experimental implementations~\cite{LLW+17,XWP+17,MMR18,LXW+18,PJO+20,GRM20}.



There are scenarios beyond the common setting of channels.
A mathematical milestone is the development of quantum superchannel (and comb) theory~\cite{CDP08a,CDP08,CDP09}
based on channel-state duality~\cite{Jam72,Cho75}.
It has been used in quantum metrology~\cite{CAP08}, computing schemes beyond circuit model~\cite{CAPV13},
and recently used in resource theory of channels~\cite{Gour19}, 
non-Markovian dynamics 
with quantum memory effects~\cite{LHW18},
and quantum algorithms~\cite{Wang21}.
Therefore, it is worthy to explore more applications of superchannels in quantum computing.

The quantum channel simulation approach developed in \cite{WBOS13,WS15,Wang16}
is based on the convex-decomposition of channels into sum of extreme ones.  
The convex-decomposition of channels relies on the convex geometry of channels
acting on a quantum system.
The nontrivial feature is that extreme channels can have ranks higher than one~\cite{Cho75,LS93,Dar04,Rus07,FL13,IC18,MS22},
making the convex-decomposition difficult.
Actually, there is a notable open problem called Ruskai's conjecture~\cite{Rus07}
which concerns the upper bound for the number of (generalized) extreme channels needed.
This has motivated mathematical investigation on the convex geometry~\cite{FL13,IC18}.
The algorithmic approach developed in \cite{WBOS13,WS15,Wang16} has numerically verified 
Ruskai's conjecture for low-dimensional cases (up to four).  
Also the circuit cost is smaller than a direct Stinespring's dilation method.
If the ancilla is recyclable and refreshable,
the ancilla can be as small as a qubit, 
which adaptively controls a sequence of operations to simulate a channel~\cite{LV01,AO08,SNA+17,ICC17}.

In this work, we extend quantum channel simulation to quantum circuit simulation of superchannels,
or quantum superchannel simulation.
It is the expansion of superchannel theory to quantum simulation and algorithm on the one hand, 
and on the other hand, it is also the extension of quantum channel simulation to superchannels.  
In particular, we develop the simulation scheme based on convex-decomposition of 
extreme superchannels.
This requires the description of extreme superchannels,
which has not been systematically studied before.
We make a parameter counting for superchannels and combs,
and present an analog conjecture with Ruskai's for superchannels.
We develop numerical optimization simulation algorithm for qubit superchannels,
and present simulation results for a few cases. 
Our results can be used to study more general superchannels
and employed for various experimental implementation.


This work contains the following parts.
In section~\ref{sec:pre}, we review the representations of quantum channels,
and the convex-decomposition algorithm of channels.
We revisit the exact decomposition of channels~\cite{RSW02,Rus07},
and argue that the decomposition is non-unique and most likely can only be done numerically.
In section~\ref{sec:comb}, we develop the representations of superchannels and extreme ones,
and also the convex-decomposition algorithm,
with numerical simulations for the case of qubit superchannels.
We also discuss the simulation of unital channels and superchannels,
and the physics of extremality (in the appendix).
We conclude in section~\ref{sec:conc}.

\section{Quantum circuit simulation of channels}
\label{sec:pre}


\subsection{Channels and extreme channels}
\label{sec:ext-rep}

In this section, we describe the theoretical framework for 
quantum circuit simulation of superchannels by reviewing the case 
of channels which employs a convex-decomposition optimization algorithm.

We focus on finite-dimensional Hilbert spaces. 
We first recall representations of quantum channels briefly. 
For a Hilbert space $\S H$, 
quantum evolution is 
in general described by CPTP maps~\cite{Kra83},
or known as quantum channels
\be \C E(\rho)=\sum_{i=1}^r K_i \rho K_i^\dagger, \ee
for $K_i$ as Kraus operators~\cite{Kra83} with $\sum_i K_i^\dagger K_i=\I$,
and $r$ as the rank of $\C E$, $\rho \in \S D(\S H)$.
Recall that the Kraus representation above is not unique, 
and the rank is the minimal one among all equivalent representations. 
As an example, unitary evolution is rank 1 with $U^\dagger U=UU^\dagger=\I$ 
for $U\in SU(d)$, with $d=\text{dim}(\C H)$.
The notable channel-state duality~\cite{Jam72,Cho75}
maps a channel $\mathcal{E}\in\mathscr{L}(\mathscr{D})$ into a quantum state
\begin{equation}\label{eq:choi}
\omega_\C E:=\mathcal{E} \otimes \mathds{1} (|\omega\rangle\langle\omega|),
\end{equation}
called Choi state $\omega_\C E\in \mathscr{L}(\mathscr{H}\otimes\mathscr{H})$, 
for $|\omega\rangle:=\frac{1}{\sqrt{d}}\sum_{i=0}^{d-1}|i,i\rangle$
known as a Bell state or ebit.
The rank $r$ is concisely characterized as the rank of the Choi state $\omega_\C E$. 
The Stinespring's dilation theorem~\cite{Sti55} guarantees that 
a channel can be described as an isometry $V$ with $V=\sum_i |i\ket K_i$,
and further dilated to a unitary $U$ with $V=U|0\ket$ as the first block column of $U$,
and $|0\ket$ as the initial ancillary state. 
We will refer to this as the quantum circuit representation of channels.




An important feature of channels is the convexity of their set. 
A convex set has extreme points, 
which cannot be written as any convex combination of others.
It is clear to see unitary operators are extreme, 
but there are more extreme channels with higher ranks.
For the set of qudit channels $\mathscr{S}_{d}$,
Choi~\cite{Cho75} proved that 
a channel $\C E$ is extreme in $\mathscr{S}_{d}$ iff $\{K_i^\dagger K_j\}$ is linearly independent
for $K_i$ as its Kraus operators. 
A consequence is that the rank of a channel necessarily satisfies $r\leq d$ in order for being extreme,
while for general channels the rank is upper bounded by $d^2$.
Define the set of rank up to $d$ channels as $\mathscr{S}^{g}_{d}$,
and such channels are called generalized extreme channels~\cite{Rus07},
or ``gen-extreme'' channels following Ref.~\cite{W20_choi}.
A gen-extreme channel that is not extreme is called a quasi-extreme channel,
which clearly can be written as a convex sum
of at least two extreme channels.


The mathematical conditions above for being extreme or gen-extreme 
are not enough for applications.
We still need to find formulas of them.
Here we review the representations of extreme channels,
including the Choi state form~\cite{WS15} and quantum circuit form~\cite{WS15,Wang16,ICK+16,ICC17,W20_choi}.


Choi state is a bipartite matrix, 
which allows a natural block-matrix form representation.
There is a concise block-matrix form of gen-extreme channels~\cite{WS15} as
\bea \omega_\C E &=&\sum_{ij} |i\ket \bra j| \otimes \C C_{ij},\\
 \C C_{ij}&:=&\C E^t(|i\ket \bra j|) = \sqrt{\C C_{i}}U_i^\dagger U_j \sqrt{\C C_{j}}\eea
for $\C C_{i}\equiv \C C_{ii} \geq 0$, and $U_i, U_j\in \text{U}(d)$.
For general channels, the unitary operators above would be replaced by contractions~\cite{WS15}.
Also a peculiar feature of gen-extreme channels is that 
the form above can be used directly to find the quantum circuit for the channel,
without solving the Kraus operators~\cite{W20_choi}.
This is generically not possible for general channels without knowing the Kraus operators.
Namely, define an isometry $V:=\sum_i |i\ket U_i \sqrt{\C C_{i}}$, then 
\be \C E^t(\rho)=\sum_{ij}\rho_{ij} \C C_{ij}=V^\dagger (\rho \otimes \I) V, \ee
for $\rho=\sum_{ij}\rho_{ij} |i\ket \bra j|$.
Now the channel is realized by a quantum circuit $W$ with 
\be \C E(\rho)=\text{tr}_a \C W (\rho\otimes |0\ket\bra 0|), \ee
for $W=\text{swap} \cdot  U^*$, 
and the swap gate, which is applied after $U^*$, is on the system and ancilla which are of the same dimension.
The unitary $U$ is defined such that $U|0\ket=V$.
The Kraus operators are derived as $K_i=\bra i|W|0\ket$.

Knowing the quantum circuit $W$ is not enough yet since we still need to decompose it 
into product of elementary gates. 
In quantum computing, we often use the universal gate set $\{\textsc{cnot}, \text{U}(2)\}$,
for the controlled-not gate $\textsc{cnot}$ and the group $\text{U}(2)$ on qubit.
Matrix decomposition techniques have been employed to design quantum circuits for channels,
including the cosine-sine decomposition (CSD) and QR decomposition~\cite{WS15,Wang16,ICK+16,ICC17}.
A qubit gate can be further decomposed as product of Hadamard gate and T gate for fault-tolerance~\cite{NC00},
but in this work we do not require this.
To quantify the complexity or cost for executing a circuit,
we refer to the circuit cost as the size of required ancilla, 
and the total number of $\textsc{cnot}$ and qubit gates in a circuit,
which scales linearly with the number of free parameters in the unitary operator for the circuit~\cite{ICK+16}.
Also for the setting of our algorithm, we mostly use the number of free parameters 
to characterize the complexity of a circuit or circuit family. 

\subsection{Convex decomposition of channels}

Given the convex body of channels, we expect a channel can be decomposed as a sum of extreme channels,
just as a state decomposed as a mixture of pure states.
However, convex decomposition of channels is difficult, 
and it is still an open problem in the field of quantum information since 2007~\cite{Rus07}.
Namely, Ruskai and coauthors conjectured that 
any channel $\mathcal E \in\mathscr{S}_{d}$ can be decomposed as
\begin{equation}\label{eq:rus}
  \mathcal E= \frac{1}{d} \sum_{i=1}^d \mathcal{E}^\emph{g}_i.
\end{equation}
The uniform probability in the sum (\ref{eq:rus})
can be relaxed to general mixing with probability $p_i\in [0,1]$ and $\sum_{i=1}^d p_i=1.$
The case for $d=2$ has been proven~\cite{RSW02},
also revised in the next section,
the method of which can be extended to the case for dimension-altering channels 
from a qudit to a qubit~\cite{Rus07},
a useful fact that will be used below for our simulation of qubit superchannels.
This conjecture is numerically supported for dimensions up to four~\cite{WS15,Wang16}.
The classical optimization algorithm aims at
\be \min_{\{ p_i, \vec{\theta}\}} D_t (\omega_\C E, \omega_\C E') \ee
for $\vec{\theta}$ as the angle parameters in $\omega_\C E'$, which is a sum of gen-extreme channels,
and $\omega_\C E$ as a random input Choi state for a channel $\C E$. 
The objective function $D_t (\omega_\C E, \omega_\C E')$ is the trace distance between states,
upper bounded by 1.
Practical simulation finds that $p_i$ can be chosen to be uniform,
which is suitable for experimental implementation~\cite{LLW+17}.

It is unlikely that there exists an exact decomposition for general channels.
The Carath\'{e}odory theorem on convex sets~\cite{GW93} provides 
the upper bound $d^4-d^2$ as the number of extreme channels,
which is too loose in this case.
As the rank of extreme channels is not fixed, 
it is hard to characterize them rank by rank. 
Ruskai's conjecture is also meaningful from a heuristic parameter-counting.
A general qudit channel contains $d^4-d^2\in O(d^4)$ parameters,
while a rank-$d$ gen-extreme channel contains $2d^2(d-1)\in O(d^3)$ parameters,
hence it is reasonable to require a mixing of $d$ gen-extreme channels.
Compared with eigenvalue decomposition of matrices,
the convex decomposition of channels can be viewed as a sort of generalized eigenvalue decomposition.
A gen-extreme channel can be mapped to an isometry $V$ as described above,
which can be further treated as a pure state. 
The difficulty of this decomposition shall be comparable with the eigenvalue 
decomposition of Choi state directly to find the set of Kraus operators.
The merit for quantum simulation is to reduce the circuit size, 
from $O(d^4)$ gates to $O(d^3)$ gates and from three qudits to two qudits,
which would benefit experimental implementation in the near future.

\subsection{Example: qubit channel decomposition}

Two decades ago, Ruskai, Szarek, and Werner~\cite{RSW02} first proved 
that a qubit channel can be decomposed into a sum of two gen-extreme qubit channels.
We will refer to it as the RSW method. 
Much later, the method based on cosine-sine decomposition (CSD) is developed~\cite{Wang16} 
for an optimization algorithm.
Within this framework, 
we emphasize that the convex decomposition is not unique 
and 
generically can only be done numerically, 
i.e., cannot be done analytically 
for arbitrary channels.

The RSW method mainly contains two ingredients:
an exact decomposition based on the Choi state
and a canonical form of extreme qubit channels based on the affine representation of channels.
Up to a pre and a post unitary gate, it shows that 
a gen-extreme qubit channel is specified by 
two Kraus operators 
\begin{equation}\label{eq:extqubF}
  F_0=\begin{pmatrix}\cos\beta &0\\0&\cos\alpha\end{pmatrix}, \quad
	F_1=\begin{pmatrix}0&\sin\alpha\\\sin\beta&0\end{pmatrix},
\end{equation}
with $\alpha$, $\beta$ as rotation angles.
Based on the block-matrix form (see Sec.~\ref{sec:ext-rep}),
any qubit channel $\mathcal{E}$ can be written as an average of two gen-extreme qubit channels
  \begin{equation}\label{eq:qubdecompcs}
    \mathcal{E}= ( \mathcal{E}_1^\emph{g} + \mathcal{E}_2^\emph{g} )/2
  \end{equation}
since a two-by-two contraction matrix $A$ can be decomposed by singular-value decomposition 
as an average of two unitary matrices 
\be A= V(U+ U^*)W/2. \ee
The two gen-extreme qubit channels in the RSW decomposition have the same diagonal blocks 
$\C C_0$ and $\C C_1$ in their Choi states.
From parameter-counting, a qubit channel needs up to 12 parameters,
but only 8 for gen-extreme ones.
From the Choi state, there are 4 parameters for $\C C_0$ and 8 parameters for a contraction $A$,
but 4 parameters for a unitary $U\in \text{U}(2)$.
This shows how the parameters split under the decomposition.
In addition, this method extends to channels that map qudit to qubit systems~\cite{Rus07}
due to the two-by-two block-matrix form of Choi states,
but cannot be extended further to qudit channels~\cite{WS15}.

With the map between Choi state and circuit for gen-extreme channels defined in Sec.~\ref{sec:ext-rep},
we find 
a more explicit form
\begin{subequations}
\be K_0=P_0 \sqrt{\C C_0^*}+ Q_1 U^*\sqrt{\C C_1^*}, \ee
\be K_1=Q_0 \sqrt{\C C_0^*}+ P_1 U^*\sqrt{\C C_1^*}, \ee
\label{eq:extqubK}
\end{subequations}
for a unitary $U\in \text{U}(2)$, $P_0=|0\ket \bra 0|$, $P_1=|1\ket \bra 1|$,
$Q_0=|0\ket \bra 1|$, $Q_1=|1\ket \bra 0|$.
Despite this,
the forms between $F$~(\ref{eq:extqubF}) and $K$~(\ref{eq:extqubK}) do not match.
If one wants to use the $F$ form, 
it still needs to find the pre $U_\text{pr}$ and post $U_\text{po}$ unitary rotations on the qubit, 
and also possibly a basis transformation $U=[u_{ij}]$ on the ancilla.
Namely, $K_i=\sum_j u_{ij} U_\text{pr} F_j U_\text{po}$,
which apparently cannot be analytically solved in general. 
However, due to the diversity of channel representations
and the rich structure of the set of channels,
it is still left open whether there could be a better analytical method 
for the convex-decomposition of generic or special types of channels. 


The theoretical analysis above is beneficial to understand 
the decomposition problem, but for the simulation algorithm 
we need to follow the algorithmic approach that we have developed~\cite{WBOS13,WS15,Wang16}. 
In order to motivate the circuit simulation of superchannels,
now we recall the CSD method which directly represents gen-extreme channels by quantum circuits~\cite{Wang16,ICC17,ICK+16}.
For the qubit case, 
we need the unitary operators $U\in \text{U}(4)$ which can be represented as
\begin{equation}\label{eq:Uqubgext1}
  U=\begin{pmatrix}     W_1 & \\ & W_2   \end{pmatrix}
  \begin{pmatrix}     C & -S \\ S & C   \end{pmatrix}
  \begin{pmatrix}     V & \\ & V_2   \end{pmatrix},
\end{equation}
for $C\equiv \text{diag}(\cos\theta_1,\cos\theta_2)$,
$S\equiv \text{diag}(\sin\theta_1,\sin\theta_2)$,
$V,V_2,W_1,W_2\in \text{U}(2)$.
Then the Kraus operators are 
\begin{equation}\label{eq:Kqubgext}
K_0=W_1 C V, \; K_1=W_2 S V.
\end{equation}
The above form~(\ref{eq:Kqubgext}) was used as an ansatz in an optimization algorithm
for the convex-decomposition of qubit channels,
which takes a desirable qubit channel as input and yields the parameters in the circuits directly~\cite{WBOS13,WS15,Wang16}.
The CSD method extends to qudit channels and we will use this for superchannels, as well.

\section{Quantum circuit simulation of superchannels}
\label{sec:comb}

\begin{figure}
    \centering
    \includegraphics[width=.55\textwidth]{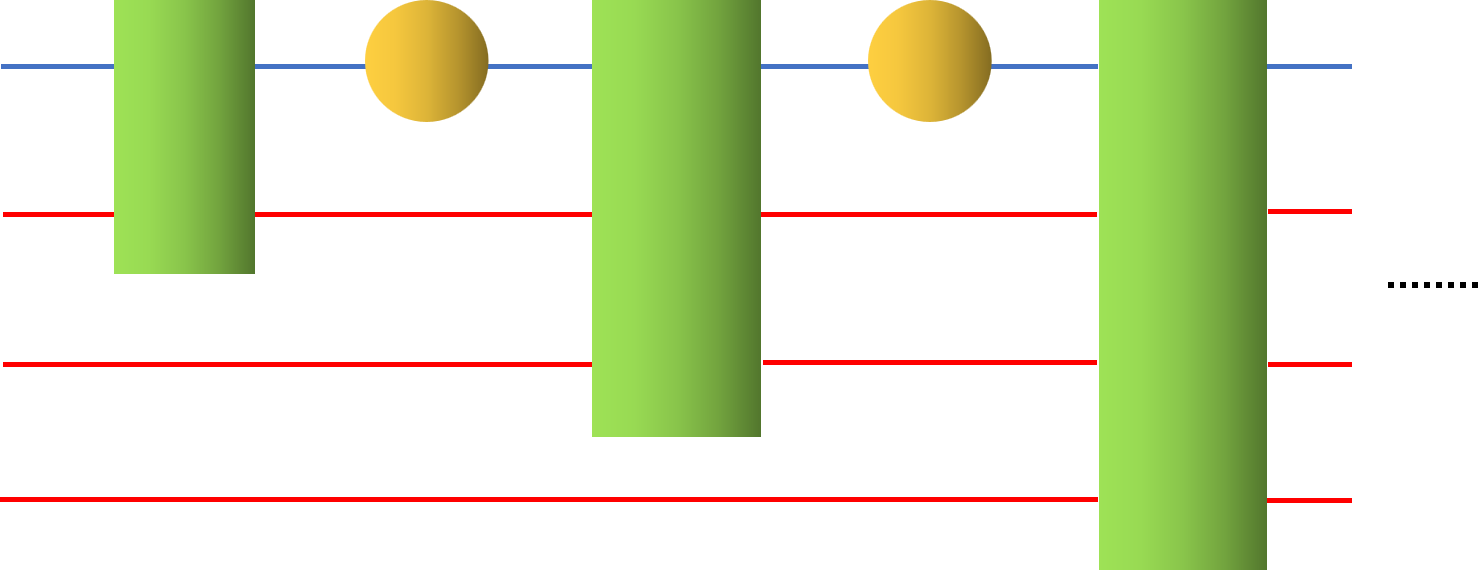}
    \caption{Schematics of quantum circuit for quantum $n$-superchannels with $n+1$ `boxes' (for unitary operators) and $n$ `circles' (for input channels).
    The figure is for a 2-superchannel.
    The top wire (blue) is for a qudit system, and all other wires (red) are for ancillas.
    The colors highlight different components of the circuit.}
    \label{fig:qcomb}
\end{figure}

\subsection{Superchannels and extreme superchannels}

In this section, we review the theoretical framework of quantum $n$-combs~\cite{CDP08a,CDP08,CDP09},
which include superchannels for $n=2$.
We will elucidate the circuit representation and present 
an analog of Ruskai's conjecture for superchannels. 
As channels can be represented as Choi states,
the mappings on Choi states are further defined as superchannels.
Similar frameworks have also been introduced~\cite{GW07,Jen11,Jen12,Jen14},
while here we employ the most well developed framework of quantum combs.
By definition, channels are 1-combs, and states are 0-combs.
However, in this work we will change the terminology slightly.
We will define $n$-superchannels as $(n+1)$-combs to simplify the notions.
So channels are 0-superchannels, the usual superchannels are 1-superchannels.
From Fig.~\ref{fig:qcomb}, $n$-superchannels will have $n$ circles for the input,
while $n+1$ unitary `boxes' as the `teeth' in the comb.

We focus on Choi state and quantum circuit representations,
while other forms such as the affine representation can also be developed.
We consider the qudit case. 
Label the input space for the channels as $\S H_0$, $\S H_2$, $\dots$, $\S H_{2n}$,
and the output space for the channels as $\S H_1$, $\S H_3$, $\dots$, $\S H_{2n+1}$.
Let an orthonormal operator basis for each space $\S L(\S H)_k$ be $\{E_{i,[k]}\}_{i\in [1,d^2]}$ with $E_{1,[k]}=\I$,
and all others traceless.
As we use $\omega_{\C E}$ as Choi state for a channel $\C E$,
we use $\omega^{(n)}$ as the Choi state of an $n$-superchannel,
which is shown~\cite{DPS11} to be the form
\bea \nonumber
\omega^{(n)}&=&\frac{1}{d^{2(n+1)}}\I_{2n+1,\dots,0} + 
\sum_{i=2}^{d^2}E_{i,[2n+1]}\otimes B_{i,[\overline{2n}]} \dots \\
&+& \I_{2n+1,2n}\otimes \sum_{i=2}^{d^2}E_{i,[2n-1]}\otimes B_{i,[\overline{2n-2}]} \dots \\ \nonumber
&+& \I_{2n+1,\dots,2}\otimes \sum_{i=2}^{d^2}E_{i,[1]}\otimes B_{i,[0]},
\eea
with arbitrary hermitian operators $B_{i,[\overline{k}]}$ acting on $\otimes_{j=0}^k\S H_j$.
The form is recursive with 
\be \omega^{(n)}=\frac{1}{d^2}\I_{2n+1,2n}\otimes \omega^{(n-1)}+\sum_{i=2}^{d^2}E_{i,[2n+1]}\otimes B_{i,[\overline{2n}]}.
\ee
Before we present the quantum circuit form, we make a parameter counting for $\omega^{(n)}$,
which has not been presented explicitly before. 
The free parameters are contained in the $B$ matrices above.
Let $y_n$ be the number of parameters for $\omega^{(n)}$, then 
\be y_n=y_{n-1}+(d^2-1)d^{4n+2}.\ee
For channels, $y_0=d^2(d^2-1)$, so we find 
\be y_n= \frac{d^2}{d^2+1}(d^{4(n+1)}-1).\ee
For superchannels $y_1=(d^4+1)d^2(d^2-1)$.
We find from this the dilated quantum circuit is to add one additional ancilla of dimension $d^2$,
see Fig.~\ref{fig:qcomb}.

Given a Kraus form of a channel $\C E=\{K_i\}$, it is changed by a superchannel $\hat{\C S}$ as 
\be \hat{\C S}(\omega_{\C E})=\sum_a S_a \omega_{\C E} S_a^\dagger\ee
with \be S_a= \sum_m K_w^{ma} \otimes K_v^m \label{eq:s}\ee 
and $\sum_a S_a^\dagger S_a=\I$.
The new channel is represented by Kraus operators
\be F_i^a= \sum_m K_w^{ma}K_i K_v^{m, \mathsf{T}},\ee
with $\sum_{ia}F_i^{a\dagger}F_i^a=\I.$
Note we put a hat on the symbol for superchannels,
and the superscript $\mathsf{T}$ is transpose.
The operators $K_v^m$ and $K_w^{ma}$ are Kraus operators,
and their meaning can be seen clearly from the quantum circuit form of a superchannel, 
see Fig.~\ref{fig:qcomb}.
We have $K_v^m=\bra m|V|0\ket$ and $K_w^{ma}=\bra m|W|a\ket$ for unitary operators $V$ and $W$.

The circuit form in Fig.~\ref{fig:qcomb} takes the channels as input directly, plugged in the circuit `board'~\cite{CDP08} at suitable locations instead of starting from the initial time. 
This can also be equivalently realized by a circuit acting on 
their Choi states as input at the initial time~\cite{Wang22},
based on the channel-state duality, in particular, 
the map between the identity operator and the ebit state. 
It is apparent that superchannels are special channels,
so they can be simulated by circuits for channels. 
However, the simulation cost might be larger. 
For a rank $r$ superchannel, the size of pre $V$ and post $W$ unitary is of size at most $rd$.
If we use a channel to simulate its action of Choi state,
then it needs a unitary dilation $U$ of size $rd^2$.
The particular form of Kraus operators $S_a$ (\ref{eq:s}) 
needs to be taken into account to obtain an optimal circuit.
For the goal of quantum simulation in this work, 
it is suitable to employ the circuit form in Fig.~\ref{fig:qcomb}. 

There is no direct relation between $U$ and the pair of $V$ and $W$.
However, if $V$ is of special type, we can obtain $U$ from the pair.
From CSD, $V$ can be written as a block form,
which is a product of multiplexers~\cite{Wang16}.
Recall that a multiplexer is a type of uniformly-controlled gates, 
which is block-diagonal in a certain basis.
Here $V$ is bipartite acting on the system and an ancilla. 
A multiplexer either has the system or the ancilla as the control.
So, if $V$ only contains one multiplexer,
then it can be shuffled to delete the ebits.
The $U$ is of the form $U=(\I \otimes W)(\I \otimes V')$,
with $V'$ obtained from $V$, and the two $\I s$ each acts on one part of the Choi state.

Furthermore, the adaptive-circuit approach developed in~\cite{LV01,AO08,SNA+17,ICC17}
reduces the ancillary cost to a single qubit for any channels,
and no qubit ancilla needed for POVM,
yet this approach does not apply to superchannels.
The reason is that the pre and post unitary operators cannot be reduced to isometries
due to the memory wire in between, 
which transfers quantum information
from the pre to the post unitary operators. 
This highlights the peculiar effect of quantum memory realizing superchannels
which proves to be useful in various tasks~\cite{CAP08,CAPV13,LHW18,Gour19,Wang21}. 

The extremality condition of $n$-superchannel has been proven by D'Ariano and coauthors~\cite{DPS11},
which is an algebriac condition that is more complicated than the case for channels.
While we find the necessary condition is that the set $\{K_i^\dagger K_j\}$ is linearly independent 
for $K_i$ as the Kraus operators for an $n$-superchannel.
This further reflects on the ancillary dimensions in the quantum circuit form.
For clarity, let us denote the ancilla dimensions by a vector,
then we find
the ancilla dimension for extreme $n$-superchannel is alternating between $d^2$ and $d$,
namely, 
\be (d^2,d,d^2,d,\dots).\ee
If $n$ is even, we need to modify the final ancilla dimension.
For example, if $n=2$, the ancilla dimension is $(d^2,d^2)$.
From the parameter counting, we find a full-rank 
gen-extreme $n$-superchannel needs parameters in the order $O(d^{3(n+1)})$.
As a result, we conjecture that
\be \omega^{(n)}= \sum_{i=1}^{d^{n+1}} p_i \omega^{(n),g}_i,\ee
which extends Ruskai's conjecture for qudit channels in Eq.~(\ref{eq:rus}).
The above conjecture is one of the central result in this work.
Below we numerically support this conjecture for qubit superchannels
with $n=1$ and $d=2$.
Also it is worthy to mention that the RSW method
does not directly extend to the case of qubit superchannels.


\subsection{Convex decomposition of superchannels}
\label{sec:2combdecom}

\begin{figure*}[t!]
    \centering
    \includegraphics[width=0.7\textwidth]{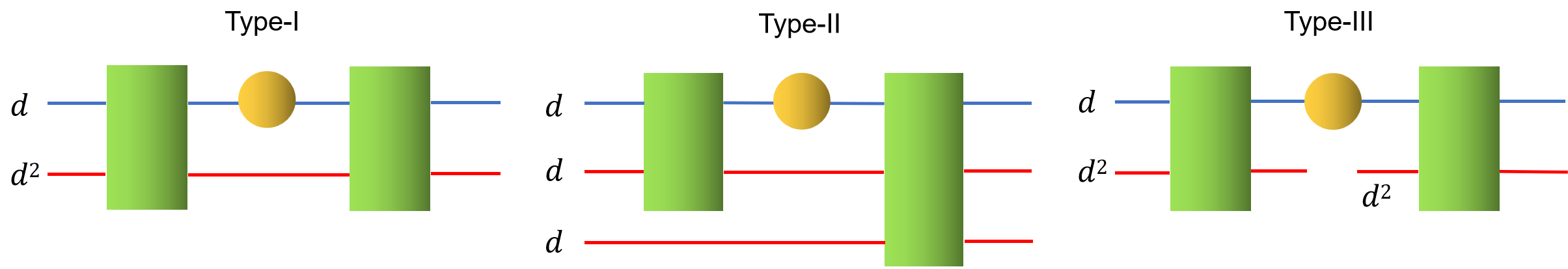}
    \caption{Three types of circuits for gen-extreme qudit superchannels.}
    \label{fig:ext-2comb}
\end{figure*}

We first propose a few quantum circuit forms for gen-extreme superchannels,
and then apply them to the qubit case. 
In the above we know that qudit superchannels contain $O(d^8)$ parameters,
while gen-extreme ones contain $O(d^6)$ parameters.
This means it needs an ancilla of dimension $d^2$ for gen-extreme superchannels,
as shown in Fig.~\ref{fig:ext-2comb}.
The post unitary $W$ contains $O(d^6)$ parameters,
while the pre unitary $V$ contains $O(d^4)$ parameters.
This is named as type-I circuit.
We also introduce type-II and type-III circuits as shown in the figure as subclasses of type-I.
The type-II circuit contains $O(d^5)$ parameters,
and type-III circuit contains $O(d^3)$ parameters.
Beyond decomposition task, these circuits can be used to generate different types of gen-extreme channels.
The type-I is general, while type-II only contains a memory ancilla of dimension $d$, 
and type-III does not have a memory wire in between, namely,
the two unitary $V$ and $W$ realize two independent channels.
As such, they can be further reduced to gen-extreme channels.
In terms of Choi states, it is a product state for type-III,
while entangled for the other two.
The block-matrix form of Choi states is not apparent for superchannels
compared with channels,
so we will not present them here.
 
For qubit superchannels, the convex-decomposition needs four gen-extreme ones of type-I.
We use CSD to represent the pre and post unitary operators $V\in \text{U}(8)$ and $W\in \text{U}(8)$.
The quantum circuit can be easily designed using iterative CSD as in \cite{WS15,Wang16,ICC17}.
A circuit form can also be found in \cite{VMS04}.
If we count the number of $\textsc{cnot}$ gates,
they scale with the number of parameters, 
and this is much larger than the number for gen-extreme qubit channels,
which is only one. 
Therefore, we see there is a big difference between channels and superchannels. 

The type-II circuits have an interesting connection with channels.
For a general qubit superchannel, 
from the viewpoint of $W$, it can be viewed as the dilation of a channel
from three-qubit to a qubit.
From Ruskai's proof~\cite{Rus07}, it can be decomposed into a sum of two gen-extreme 
channels from three-qubit to a qubit, 
which only needs a qubit ancilla. 
Namely, the ancillary dimension vector is $(4,2)$.
Denote the set of these rank-8 channels as $\S S^8$,
and they are not gen-extreme qubit superchannels.
This means they can be further decomposed into type-I or type-II circuits,
which have ancillary dimension vector $(4,0)$ and $(2,2)$, respectively.
However, we expect no exact decomposition of circuits of the form in $\S S^8$.
Therefore, in our numerical simulation algorithm 
we take an anonymous approach that employs the most general type-I circuits 
to represent gen-extreme qubit superchannels,
and take arbitrary qubit superchannels as input.

\subsection{Unital channels and superchannels}
\label{sec:unital}

Besides the generic case, 
it is also important to see if there could be simpler scheme for special types of input superchannels.
For the important class of unital ones, we find, different from channels, it is mostly not the case
for superchannels. 
We first recall the case of unital channels.
A channel is unital iff it preserves the identity operator $\mathds{1}$.
Unital channels form a convex set, denoted as $\S S^\textsc{u}_d$, and different from extreme channels,
a product of unital channels is still unital.
The extreme points for unital channels have also been characterized~\cite{LS93},
which requires that $\{K_i^\dagger K_j\bigoplus K_j K_i^\dagger \}$ is linearly independent.
However, this condition leads to an upper bound of rank for gen-extreme unital channels 
even larger than the usual case.
No circuit form for extreme unital channels has been known. 
Rather than using extreme unital channels, 
it turns out there is another scheme that is more suitable for simulation.  
It is proved that~\cite{MW09} a channel $\mathcal{E}$ is unital iff it is an affine combination of unitary channels
 \be \mathcal{E}=\sum_i a_i \mathcal{U}_i, \ee 
 with $\sum_i a_i=1$, and $a_i\in \mathbb{R}$.
This is a quantum extension of the Birkhoff's theorem,
which states that doubly stochastic matrices are in the convex hull of permutations.
However, extreme unital channels can have rank higher than 1.
Mixing-unitary (MU) channels only forms a subset of unital channels. 
Namely, we have 
\be \S S^\textsc{mu}_d \subset \S S^\textsc{u}_d \subset \S S_d.\ee
For qubit channels, $\S S^\textsc{mu}_2 = \S S^\textsc{u}_2$. 
For qutrit, there is a simple example of unital but not MU channel,
whose Kraus operators are from the generators of SO(3) in the spin-1 representation~\cite{LS93}. 

Therefore, all unital channels can be simulated by sampling unitary operators,
with possible negative probability. 
There is no need of a quantum ancilla,
which means all initial quantum information in the initial state $\rho$ is preserved. 
This implies that unital channel preserves or does not contain quantum information. 
So it is reasonable that unital channels preserve $\I$, i.e., the junk state.

For a non-unital channel $\C E$, it changes the input $\I$, namely, 
it can create quantum information $\C E(\I)$.
It requires an ancilla, whose final state is 
\be \sigma= \sum_{ij}\text{tr}(K_j^\dagger K_i) |i\ket\bra j|. \label{eq:sigma}\ee
It is a Gram matrix of the set of vectors $\{K_i\otimes\I|\omega\ket\}$.
The Gram matrix is of full rank if $\{K_i\}$ is linearly independent.
Both $\C E(\I)$ and $\sigma$ can be used to quantify the non-unital feature of a channel.
It is interesting to observe that, physically,
being unital is quite distinct from being extreme. 
The formula~(\ref{eq:sigma}) above is better understood from its complementary channel,
discussed in the Appendix~\ref{sec:ext}. 

Now we study unital superchannels.
It turns out the structure is more complicated,
and different generalizations of unital channels can be defined~\cite{Gour19}.
They can be completely specified by conditions on the Choi states of superchannels which take the form
\be \omega^{(1)}=\frac{1}{d^2}\I_{3,2}\otimes \omega^{(0)}+\sum_{i=2}^{d^2}E_{i,[3]}\otimes B_{i,[\overline{2}]}.
\ee
For simplicity, we use $R$ as the symbol for a superchannel 
and $R_{[i]}$ to denote its marginal on the space $i$.
A set of identity-preserving (IP) superchannels is defined with 
\be R_{[0,1,3]}=\I_3\otimes R_{[0,1]}, \ee
which is denoted as condition (a).
A set of doubly-stochastic superchannels is defined with (a) and also 
\be R_{[1,3]}=\I_{1,3}, \ee
which is condition (b) and requires $V$ to be the dilation of a unital channel.
A set of unital-channel preserving superchannels is defined with (b) and also 
\be R_{[1,2,3]}=\I_{1}\otimes R_{[2,3]}, \ee
which is denoted as condition (c).
A set of MU superchannels, which is the smallest amongst here,
is defined when $V$ and $W$ are unitaries with a common source of random bits.

Different from channels, there is a memory wire between $V$ and $W$ to realize superchannels,
leading to the various cases above. 
For unital and even extreme unital superchannels,
the memory wire cannot be replaced by classical bits in general.
The condition for being extreme for each type of the unital superchannels presented above 
is unknown. As for the case of unital channels, 
we expect such conditions would not help directly for their circuit representations. 
However, we can \emph{a priori} define a set of affine-mixing-unitary (AMU) superchannels as an extension of MU superchannels, with $V$ and $W$ replaced by affine-mixing of unitaries,
possibly with different sources of mixing.
The AMU set apparently belongs to the IP set above,
therefore does not capture all extreme unital superchannels. 
At present, it is not clear if the quantum Birkhoff's theorem~\cite{MW09} can be extended to unital superchannels.
For the purpose of quantum simulation, 
the wise strategy is to take the input superchannel anonymously,
no matter it is unital or not,
and then use optimization to find the convex sum of gen-extreme superchannels.

\subsection{Convex decomposition algorithm of qubit superchannels}
\label{sec:alg}

\begin{table}[]
    \centering
    \begin{tabular}{|c|c|c|}\hline
       Decomposition  &  Parameters & Precision\\ \hline
       $\hat{\C S}   \ra \sum_{i=1}^2 \hat{\C S}_i^8$  & 656 & $10^{-3}\sim 10^{-4}$ \\ \hline
       $\hat{\C S}   \ra \sum_{i=1}^4 \hat{\C S}_i^g$  & 512 & $10^{-3}\sim 10^{-4}$ \\ \hline
       $\hat{\C S}^8 \ra \sum_{i=1}^2 \hat{\C S}_i^g$  & 256 & $10^{-2}$ \\ \hline
       $\hat{\C S}^8 \ra \sum_{i=1}^4 \hat{\C S}_i^g$  & 512 & $10^{-3}\sim 10^{-4}$ \\ \hline
    \end{tabular}
    \caption{Numerical simulation results. The rows are four different decomposition tasks for more than 20 random input instances for each case. The columns show the number of parameters and simulation precision.}
    \label{tab:sim}
\end{table}

In this section, we present the numerical simulation results 
for the convex-decomposition of qubit superchannels.
The simulation is summarized in Table~\ref{tab:sim}.
Note the circuit cost is implicit based on the CSD formula 
and the number of parameters. 
For clarity, denote the set of qubit superchannels, gen-extreme, and
rank-8 ones as $\S S$, $\S S^g$, and $\S S^8$, respectively,
and elements of them as $\hat{\C S}$, $\hat{\C S}^g$, and $\hat{\C S}^8$, respectively.
A random input is generated by a pair of Haar-random unitary operators for a superchannel.
For instance,
a random $\hat{\C S}$ is generated by Haar-random unitary $V\in \text{U}(8)$ and $W\in \text{U}(32)$.
Then the Kraus operators $\{S_a\}$ for the superchannel is obtained from Eq.~(\ref{eq:s})
with $K_v^m=\bra m|V|0\ket$ and $K_w^{ma}=\bra m|W|a\ket$ for $m\in [1,4]$, $a\in [1,16]$.
The input Choi state $\omega^{(1)}$, which is a 16$\times$16 matrix, is then obtained from $\{S_a\}$.
The optimization parameters are those rotations angles in the circuits for 
gen-extreme or rank-8 superchannels,
expressed using the CSD formula. 
The optimization uses an optimization algorithm from Matlab,
which is to perform a few local optimization first, then compare and find the best local one,
and then generate a few new local samples, and iterate such a process~\cite{WBOS13,WS15,Wang16}.

We have simulated four decomposition tasks in our simulation as shown in the table.
We pick uniform probability variables.
For each task we simulated more than 20 random input instances 
following the standard routine in \cite{WBOS13,WS15,Wang16}.
The first row is for the decomposition $\hat{\C S}=(\hat{\C S}^8_1+\hat{\C S}^8_2)/2$,
which can be exactly decomposed according to Ruskai's result~\cite{Rus07}.
Our simulation contains 656 parameters, coming from the CSD representation of the circuits.
The simulation precision, i.e., the trace distance on Choi states, 
is of the order $10^{-3}\sim 10^{-4}$.
This reveals the validity of our simulation program 
which is executed on a normal desktop computer. 
We believe that better precision can be obtained with more computational resources.
The second row is for the decomposition $\hat{\C S}=(\hat{\C S}^g_1+\hat{\C S}^g_2+\hat{\C S}^g_3+\hat{\C S}^g_4)/4$,
which uses 512 parameters and yields precision in the same order as the first row.
This shows that our scheme is good enough for practical simulation of qubit superchannels,
hence can be applied in experimental tasks which often involve noisy quantum gates.
For instance, current noisy quantum devices~\cite{LJL+10,BHL+16,GQ21,BDD+22} can achieve gate infidelity 
about of the order $10^{-3}$, often with single-qubit gate fidelity 
higher than the two-qubit one. 
With tens of gates in a circuit, the implementation error would 
dominant over the error from the numerical decomposition.

The third and forth rows in the table takes a random $\hat{\C S}^8$ as the input,
and the goal is to check if it can be decomposed as a sum of two $\hat{\C S}^g$ as discussed in
section~\ref{sec:2combdecom}.
The simulation strongly suggests that this cannot be done 
numerically via our algorithm,
and instead, a sum of four $\hat{\C S}^g$ is needed. 
Although the reverse direction, namely, 
a sum of two $\hat{\C S}^g$ can lead to rank-8 superchannels,
this apparently only occupies a small portion of the whole set $\hat{\C S}^8$. 
Physically, this means that the last qubit ancilla for the post unitary operator in $\hat{\C S}^8$
cannot be replaced by a random bit in general,
and at least two random bits are needed.
More generally, a low-rank (between 4 and 16) superchannel does not necessarily require fewer 
gen-extreme superchannels in a convex decomposition,
as confirmed by the numerical simulation for both channels~\cite{WS15,Wang16} and superchannels.
This verifies that the rank of a channel, or completely positive map, is a `global' feature of it, depending on its set of Kraus operators instead of individual ones. 
It also shows that it is not so straightforward to use the rank as a parameter
to simplify our numerical decomposition. 
Instead, our numerical algorithm is effective to decompose arbitrary input 
superchannels regardless of their ranks. 

\section{Conclusion}\label{sec:conc}

In this work, we developed the quantum circuit simulation algorithm
of superchannels. 
It contains a classical optimization algorithm that designs the 
quantum circuits for the simulation of superchannels. 
This serves as a first study of superchannels and extreme superchannels 
from the viewpoint of quantum simulation and algorithm,
laying the foundation for the near-term experimental implementation. 
Our algorithm applies to generic superchannels,
and the convex-decomposition in terms of extreme ones 
proves to be effective, and can reduce the circuit cost
compared with a direct dilation approach.
The classical circuit-design algorithm may become intractable 
when the size of input channels gets larger, 
and it remains to see if better algorithms can be developed for these cases 
or cases of particular forms of input. 
We also studied the roles of special features or conditions in our algorithm,
including being unital, extreme, and the rank of input superchannels. 
More specific ones can be studied case by case.
For instance, there are superchannels that do not need a quantum ancilla, 
such as random-unitary superchannels,
Pauli or depolarizing superchannels.
The unitary twirling operation is such an example,
which plays central roles for random benchmarking and error mitigation technique.
Other analog of channels can be defined for superchannels,
such as dephasing~\cite{PKS+21} and entanglement-breaking ones~\cite{CC20}.
Their circuit simulations are quite straightforward given their concise forms.

We have seen that a key feature of superchannels
is the memory (between the pre and post unitary operators).
This feature actually has already been employed, 
even before the emergence of superchannels,
for quantum teleportation and entanglement-assisted quantum communication~\cite{BSS+99}.
Recently, we also pointed out the potential of using superchannels
to design quantum superalgorithms and applications in
quantum von Neumann architecture~\cite{Wang21,Wang22}. 
Given the growing importance of superchannels 
for quantum information science,
it is therefore crucial to develop closer physical and algorithmic study of them.
We believe our study can prompt the development of 
superchannels and their applications.

\section*{Acknowledgements}

This work is funded by
the National Natural Science Foundation of China under Grants
12047503 and 12105343.

\appendix

\section{Physics of extremality}
\label{sec:ext}

We emphasized that unital channels do not create nontrivial states from $\I$.
Here we pursue the physical meaning of being extreme.
So far we merely know that extreme points do not allow classical mixing,
and the linearly independence condition is quite abstract.
Below we study them from the viewpoint of complementary channels. 

The complementary channel $\C E^c$ of a channel $\C E$ is defined as
\be \C E^c(\rho)= \sum_{ij} \text{tr}(K_j^\dagger K_i\rho) |i\ket\bra j|.\ee
If we treat $\rho$ as the original quantum information,
and it is in general impossible to recover it if we know $\C E(\rho)$ and $\C E$.
Only special types of channels can be recovered,
and this actually leads to the error-correction condition $\text{tr}(K_j^\dagger K_i\rho)=c_{ij}$~\cite{KL97}.
However, if the final ancilla state $\sigma_f=\C E^c(\rho)$ is known, 
then, together with the channel $\C E$, they can be 
used to recover the original state $\rho$ if $\C E$ is a full-rank extreme channel
since $\{K_i K_j^\dagger\}$ is a basis.
That is, the ancilla aims to obtain the quantum information in $\rho$ through a full-rank extreme channel.
If the extreme channel is not of full rank,
then only a partial information of $\rho$ can be recovered.
This agrees with an early scheme of using qubit extreme channels as an optimal asymmetric cloning machine~\cite{NG99},
and also inspires potential extension of error-correction theory 
using extreme channels that shall be investigated separately.

There are extreme channels that are unital.
These channels can be simulated by affine-mixing of unitaries,
hence do not require a quantum ancilla.
Therefore, we identify non-unital extreme (NUE) channels as \emph{genuinely quantum} 
since they cannot be further decomposed. 
The amplitude-damping channel is such an example,
which under composition will map any initial state into a fixed pure state,
the ground state with the lowest energy.
It has already been recognized as a resource for dissipative quantum state engineering and computing~\cite{VWC09}.



We expect that our understanding of extreme channels,
especially non-unital ones,
also applies to extreme superchannels.
The complementary superchannel acts as 
\be \hat{\C S}^c(\omega_{\C E})=\sum_{ab}\text{tr}(S^{b\dagger}S^a \omega_{\C E})|a\ket\bra b|\ee
which extends the case for channels. 
If the set $\{S_b^\dagger S_a\}$ is linearly independent and full-rank, 
then the original channel $\C E$ can be recovered given the state $\hat{\C S}^c(\omega_{\C E})$ and the set $\{S_a\}$. 
An additional feature is the memory wire between the pre $V$ and post $W$,
which forbids the reduction of a superchannel into two independent channels.






\bibliography{ext}{}
\bibliographystyle{ieeetr}


\end{document}